\documentclass{article}
\usepackage{spconf,amsmath,graphicx}
\usepackage{xcolor,multirow,cite,booktabs,bm,amssymb,url}
\usepackage[labelsep=period]{caption}
\usepackage[subrefformat=parens]{subcaption}

\newcommand{\train}[0]{\text{(train)}}
\newcommand{\target}[0]{\text{(target)}}

\newcommand{\R}[0]{\mathbb{R}}
\newcommand{\C}[0]{\mathbb{C}}
\newcommand{\Rnn}{\mathbb{R}_{\geq 0}}
\newcommand{\re}{\text{(re)}}
\newcommand{\im}{\text{(im)}}
\newcommand{\pgsec}[1]{\smallskip \noindent \textbf{#1}:}
\newcommand{\argmin}{\mathop{\rm argmin}\limits}

\allowdisplaybreaks
\title{Sampling-Frequency-Independent Universal Sound Separation}
\name{Tomohiko Nakamura$^\dagger$ and\: Kohei Yatabe$^\ddagger$\thanks{This work was supported by JSPS KAKENHI under Grant JP23H03418 and JST ACT-X under Grant JPMJAX210G.}}
\address{
  $^\dagger$National Institute of Advanced Industrial Science and Technology (AIST), Tokyo 135-0064, Japan\\
  $^\ddagger$Tokyo University of Agriculture and Technology (TUAT), Tokyo 184-8588, Japan
}

\begin{document}
\ninept
\maketitle
\begin{abstract}
      This paper proposes a universal sound separation (USS) method capable of handling untrained sampling frequencies (SFs). The USS aims at separating arbitrary sources of different types and can be the key technique to realize a source separator that can be universally used as a preprocessor for any downstream tasks. To realize a universal source separator, there are two essential properties: universalities with respect to source types and recording conditions. The former property has been studied in the USS literature, which has greatly increased the number of source types that can be handled by a single neural network. However, the latter property (e.g., SF) has received less attention despite its necessity. Since the SF varies widely depending on the downstream tasks, the universal source separator must handle a wide variety of SFs. In this paper, to encompass the two properties, we propose an SF-independent (SFI) extension of a computationally efficient USS network, SuDoRM-RF. The proposed network uses our previously proposed SFI convolutional layers, which can handle various SFs by generating convolutional kernels in accordance with an input SF. Experiments show that signal resampling can degrade the USS performance and the proposed method works more consistently than signal-resampling-based methods for various SFs.
\end{abstract}
\begin{keywords}
Universal sound separation, sampling-frequency-independent convolutional layer, deep neural networks
\end{keywords}
\section{Introduction}
\label{sec:intro}
Audio source separation is a technique of separating concurrent sources from their mixture and can be used for preprocessing of various audio signal processing tasks.
Its performance has been greatly improved by the introduction of a deep neural network (DNN) \cite{mdx2021,dns2023}.
It has also expanded the range of sources that can be handled by a single source separator \cite{Kavalerov2019WASPAA,Wisdom2021ICASSP}.
These advances have opened the door to achieving one of the ultimate developmental goals in audio source separation: the realization of a source separator that can be used universally as a preprocessor for any downstream tasks.

To realize such a universal source separator, universality with respect to source types is crucial.
In usual audio source separation tasks, the target source types are specified in advance: for example, different musical instrument sounds in music source separation \cite{Takahashi2018IWAENC,Stoter2019JOSS,Liu2020Interspeech,Hennequin2020JOSS,Samuel2020ICASSP,Takahashi2021CVPR,Nakamura2021IEEEACMTASLP,Defossez2021MDX,Saito2022IEEEACMTASLP}, voices of different speakers in speech separation \cite{Hershey2016ICASSP,Luo2019TASLP,Luo2020ICASSP,Chen2020Interspeech,Zeghidour2021IEEEACMTASLP,Koizumi2021WASPAA,Takeuchi2020ICASSP,Ditter2020ICASSP}, and singing voices of different singers in vocal ensemble separation \cite{Sarkar2021Interspeech,TNakamura202306ICASSP}.
Different from these domain-specific tasks, a universal sound separation (USS) aims at separating arbitrary sounds of different types \cite{Kavalerov2019WASPAA,Wisdom2021ICASSP}.
That is, its purpose is to acquire the universality with respect to source.
Recent studies developed USS methods capable of handling weakly labeled data that contain labels of source types in each mixture but do not contain the source signals of the mixture \cite{Pishdadian2020IEEEACM_TASLP,Chen2022AAAI,Kong2023arXiv}.
These advances in USS have increased the variety of source types that can be handled by a single network.

Another important property to realize the universal source separator is universality with respect to recording conditions.
Despite its necessity, it has received less attention than the universality to source types.
Sampling frequency (SF) is one of the essential recording conditions.
The usable SFs depend on the specification of the recording devices.
They also depend on the acoustic conditions of target tasks.
Thus, the universal source separator must be able to handle various SFs with a single neural network.
However, conventional audio source separation methods (including the USS methods) commonly assume that the SF is the same in the training and test stages.
Owing to this assumption, they cannot directly handle untrained SFs and require additional preprocessing such as signal resampling.
Furthermore, we previously found that signal resampling can degrade the separation performance in music source separation \cite{Saito2022IEEEACMTASLP}.
In fact, this degradation due to signal resampling can also occur in USS, as we will show later in Section~\ref{sec:exp}.
Thus, we should explore another way to develop a single network that encompass the two properties.

\begin{figure*}[t]
  \centering
  \hspace{0.3ex}
  \begin{minipage}{1.1\columnwidth}
      \begin{center}
        \subfloat[Proposed SFI version of SuDoRM-RF]{
              \includegraphics[width=1.04\columnwidth,clip]{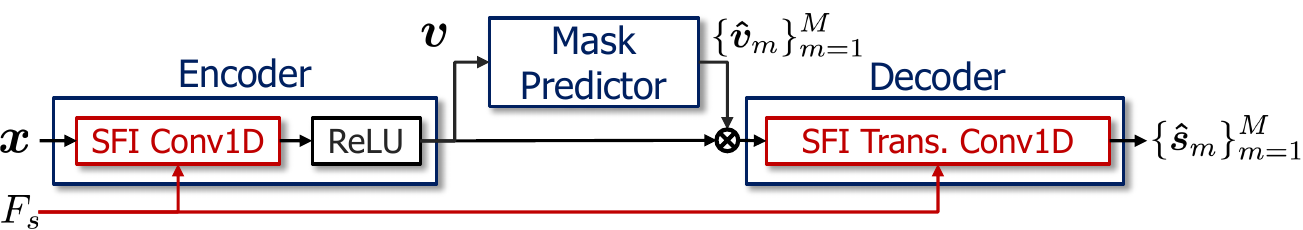}
              \label{fig:proposed}
        }
        \\
        \subfloat[SuDoRM-RF]{
              \includegraphics[width=1.04\columnwidth,clip]{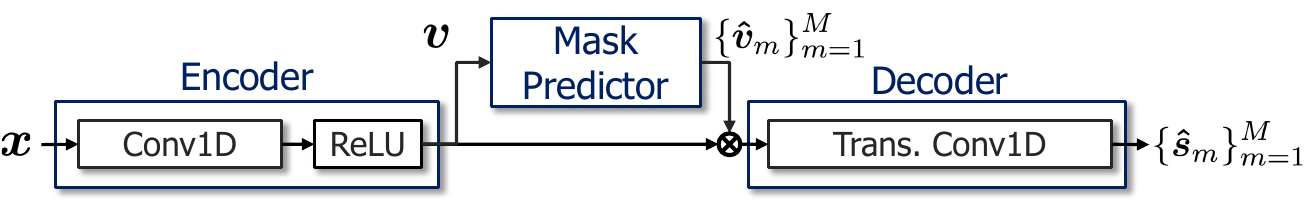}
              \label{fig:original}
        }
      \end{center}
  \end{minipage}
  \hfill
  \begin{minipage}{0.85\columnwidth}
      \begin{center}
            \subfloat[Mask predictor]{
              \includegraphics[width=0.83\columnwidth,clip]{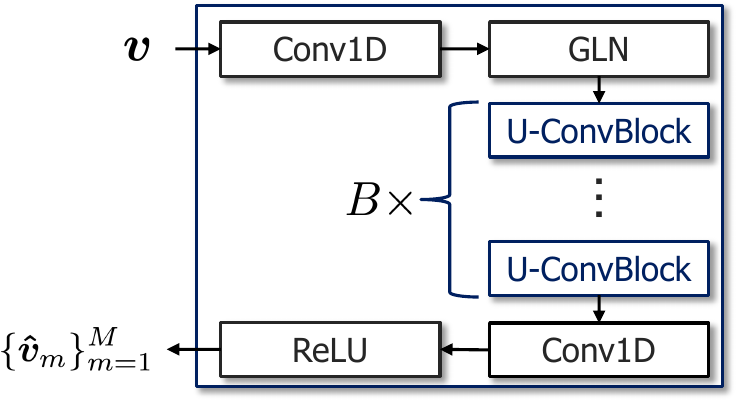}
              \label{fig:separator}
            }
      \end{center}
  \end{minipage}
  \hspace{0.3ex}
  \caption{Network architectures of \subref{fig:proposed} proposed SFI version of SuDoRM-RF, (b) original SuDoRM-RF, and (c) mask predictor. "Conv1D", "ReLU", "Trans. Conv1D", and "GLN" denote one-dimensional (1D) convolutional layer, rectified linear unit, 1D transposed convolutional layer, and global layer normalization layer respectively. See Sections~\ref{sec:sudormrf} and \ref{sec:proposed} for other variables and modules in these figures.}
  \label{fig:overview}
\end{figure*}

In this paper, we propose a USS method capable of handling various SFs (Fig.~\ref{fig:overview}\subref{fig:proposed}).
We apply our previously proposed SFI convolutional layer \cite{Saito2022IEEEACMTASLP} to SuDoRM-RF \cite{Tzinis2022JSPS}, one of the state-of-the-art USS networks.
The SFI layer can generate the weights of a usual convolutional layer in accordance with the input SF, which enables the network to handle various SFs (including untrained SFs).
A usual convolutional layer can be replaced with the SFI convolutional layer; thus, we can extend SuDoRM-RF to be universal for various SFs without losing the universality with respect to source types.
Note that the effectiveness of the SFI layer was demonstrated only for music source separation \cite{Saito2022IEEEACMTASLP}.
This paper is the first to substantiate that signal resampling can degrade the USS performance and the SFI layer is more effective than signal resampling for handling various SFs in USS.
We believe that the proposed method paves the way to realize a source separator that encompasses the universalities with respect to source types and recording conditions.

\section{SuDoRM-RF: A State-of-the-Art USS Network} \label{sec:sudormrf}
SuDoRM-RF\cite{Tzinis2022JSPS} is one of the state-of-the-art networks for USS.
In this section, we briefly show the network architecture of SuDoRM-RF and the loss function for handling a variable number of sources.

\subsection{Network Architecture} \label{sec:sudormrf_network}
Fig.~\ref{fig:overview}\subref{fig:proposed} shows the architecture of SuDoRM-RF\footnote{
      We adopted the architecture named \emph{SuDoRM-RF++ 1.0x GC} in \cite{Tzinis2022JSPS}. It achieved the state-of-the-art performance with greatly reduced parameters.
}.
It is based on the widely used framework\cite{Luo2019TASLP}, which combines trainable analysis and synthesis filterbanks with a mask predictor.
The analysis and synthesis filterbanks are called the encoder and decoder, respectively.

Let $\R$ and $\Rnn$ be the sets of the real and nonnegative numbers, respectively.
The encoder consists of a $1\times C$ 1D convolutional layer and a rectified linear unit nonlinearity.
It converts $\bm{x}\in\R^{L}$ into the pseudo time-frequency representation $\bm{v}\in\Rnn^{C\times T}$, where $L$ is the signal length.
The number of frames $T$ is given as $T=\lfloor (L+2P-K)/S+1\rfloor$, where the $K,S,$ and $P$ are the kernel size, stride, and padding of the convolutional layer in the encoder.

The mask predictor\footnote{
      In \cite{Tzinis2022JSPS}, the architecture that directly predicts the pseudo time-frequency representations of all sources was proposed to improve the separation performance.
      However, we experimentally observed that it made the training numerically unstable.
      Thus, we adopted the architecture that predicts the masks for the pseudo time-frequency representation.
} transforms $\bm{v}$ to the $M$ masks $\bm{\hat{v}}_m\in\Rnn^{C\times T}$, where $m$ is the output source index.
$M$ denotes the number of the output signals of the network and may be different from the number of sources present in the input mixture $N(\leq M)$.
Fig.~\ref{fig:overview}\subref{fig:separator} shows the architecture of the mask predictor.
The first and last convolutional layers have a kernel size of $1$ and a stride of $1$.
The main module of the mask predictor is the stack of $B$ U-ConvBlocks.
Each U-ConvBlock has the U-Net architecture of five levels and processes the input feature in multiple time resolutions by successive downsampling and upsampling.
This characteristic can capture long-term temporal dependencies without significantly increasing the number of parameters.
See \cite{Tzinis2022JSPS} for the details of U-ConvBlock.

The decoder is the $CM\times M$ 1D transposed convolutional layer with a kernel size of $K$ and a stride of $S$.
After concatenating $\{\bm{v}\odot\bm{\hat{v}}_m\}_{m}$ along the channel axis, the decoder converts it into the output signals $\bm{\hat{s}}_m\in\R^{L}$, where $\odot$ is the elementwise multiplication.

\subsection{Loss Function} \label{sec:sudormrf_loss}
In a practical situation, the number of sources $N$ is usually unknown and may vary mixture by mixture.
SuDoRM-RF can handle a variable number of sources up to $M$ by using a loss function similar to \cite{Wisdom2021ICASSP}.
Let $\mathcal{P}$ be the set of all assignments between the output and groundtruth signals, $p$ be the element of $\mathcal{P}$, and $p(n)$ be the output channel index assigned to source $n$ under assignment $p$.
The loss function is given as
\begin{align}
      \mathcal{L}=
      &
      \begin{cases}
      \min_{p\in\mathcal{P}}(\mathcal{L}_{1,p}+\mathcal{L}_{2,p}) & (M>N) \\
      \min_{p\in\mathcal{P}} \mathcal{L}_{1,p} & (M=N)
      \end{cases},
      \label{eq:loss} \\
      \mathcal{L}_{1,p}=
      &
      \dfrac{1}{N}\sum_{n=1}^{N}10\log_{10}\left(\dfrac{d_{n,p(n)}+\epsilon}{\|\bm{s}_n\|^2+\epsilon}\right)
      \\
      \mathcal{L}_{2,p}=&
      \dfrac{1}{M-N}\sum_{n=N+1}^{M}10\log_{10}(d_{n,p(n)} +\tau\|\bm{x}\|^2+\epsilon),
      \\
      d_{n,p(n)}&=\|\bm{s}_n-\bm{\hat{s}}_{p(n)}\|^2,
\end{align}
where $\bm{s}_n\in\R^{L}$ is the groundtruth signal of source $n$, $\epsilon$ and $\tau$ are small values to avoid zero divisions.
$\mathcal{L}_{1,p}$ is the negative average source-to-noise ratio (SNR) for the output signals under assignment $p$, which induces the output signals to match the groundtruth signals.
$\mathcal{L}_{2,p}$ is the loss function for the output signals that are not assigned to any groundtruth sources.
It induces the unassigned output signals to zero vectors.

\section{Proposed Method} \label{sec:proposed}
We propose an SFI extension of SuDoRM-RF by introducing our previously proposed SFI layers.
In this section, we review the SFI convolutional layer and apply it to SuDoRM-RF.

\subsection{SFI Convolutional Layer \cite{Saito2022IEEEACMTASLP}} \label{sec:sfi_layer}
The SFI convolutional layer is an extension of a usual convolutional layer.
Its idea is based on the fact that the weights of the usual convolutional layer can be interpreted as a collection of digital filters.
This interpretation reveals that the weights inherently depend on the SF and must be constructed with respect to the input SF.
To overcome this problem, we focused on an analog-to-digital filter conversion, whereby a digital filter is designed from an analog filter.
Since analog filters are SFI, we can use them as archtypes of the weights for all SFs and use this conversion to generate the weights in accordance with an input SF.
We call these archtypes the latent analog filters.

\begin{figure}[t]
  \centering
  \includegraphics[width=0.88\columnwidth,clip]{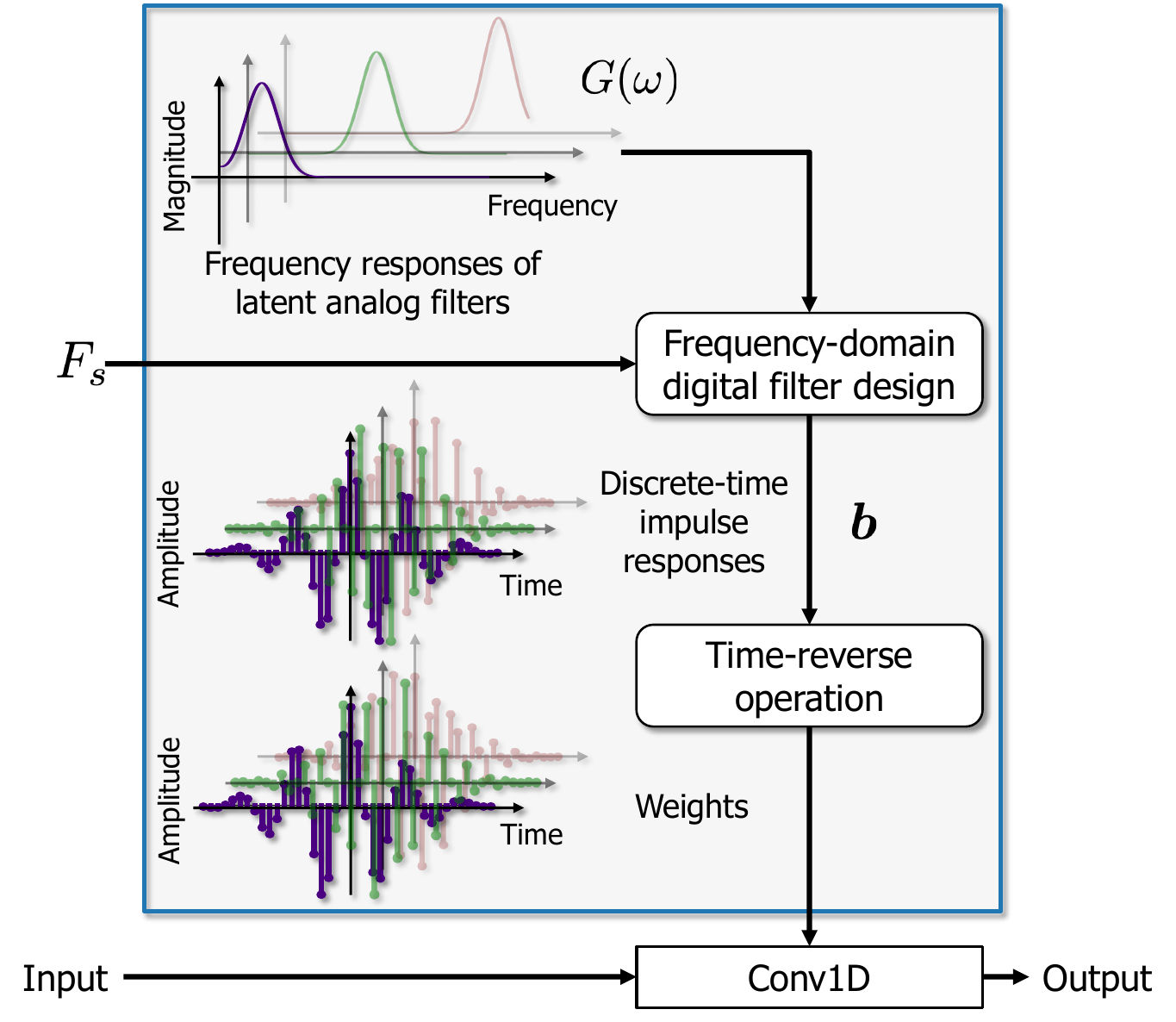}
  \caption{
        Architecture of SFI convolutional layer using frequency-domain digital filter design (adapted from Fig.~1(b) in \cite{Saito2022IEEEACMTASLP}).
  }
  \label{fig:sfi_layer}
\end{figure}

Fig.~\ref{fig:sfi_layer} shows the SFI convolutional layer using the frequency-domain filter design method.
The latent analog filters are given as continuous frequency responses for all input and output channel pairs.
Since these responses are processed in the same manner, we omit the indices of the input and output channel pair.
Let $G(\omega;\theta)$ be the continuous-time frequency response, where $\omega\in\R$ is the (unnormalized) angular frequency and $\theta$ is the set of the parameters of the latent analog filters.
Given the input SF $F_s$, the SFI layer computes the discrete-time impulse response $\bm{b}\in\R^K$ that its discrete-time Fourier transformation approximates $G(\omega;\theta)$ for $\omega\in[0,\pi F_s]$ in the least-squares sense:
\begin{equation}
      \bm{b}=\argmin_{\bm{b'}\in\R^{K}}\|\bm{G}-\bm{D}\bm{b'}\|^2, \label{eq:least_square_prob}
\end{equation}
where $\bm{G}=[G(\omega_1;\theta),\ldots,G(\omega_{I};\theta)]^\top\in\C^{I}$.
The sampled angular frequency $\omega_{i}$ is given as $\omega_i=\pi F_s i /(I-1)$, where $i=1,\ldots,I$ is the index of the sampled angular frequency.
$\bm{D}$ is the $I\times K$ matrix and its $(i,k)$th element is $e^{\j \omega_i(k-K/2)/F_s}$, where $\j$ is the imaginary unit.
This problem can be solved analytically:
\begin{equation}
      \bm{b} =
      \begin{bmatrix}
            \bm{D}^{\re} \\
            \bm{D}^{\im} 
      \end{bmatrix}^{\dagger}
      \begin{bmatrix}
            \bm{G}^{\re} \\
            \bm{G}^{\im} \\
      \end{bmatrix},
\end{equation}
where the superscript $\dagger,\re,$ and $\im$ denote the Moore--Penrose pseudo inverse of a matrix, the real part of a matrix or vector, and the imaginary part of a matrix or vector, respectively.
By reversing $\bm{b}$ of all input and output channel pairs in time, we use them as the weights of the usual convolutional layer.
Owing to this weight generation, the SFI layer can generate consistent weights for various SFs.

In the test stage, we only need to generate the weights once for each $F_s$ because the weight generation depends on $F_s$ and $G(\omega)$.
Thus, the SFI layer does not increase the computational cost except for the first weight generation.
In addition, we can construct an SFI version of a transposed convolutional layer (SFI transposed convolutional layer) by replacing the usual convolutional layer with the usual transposed convolutional layer in the SFI convolutional layer.

\subsection{Application of SFI Layers to SuDoRM-RF} \label{sec:application}
We can apply the SFI layers to SuDoRM-RF in a similar manner to the SFI network for music source separation \cite{Saito2022IEEEACMTASLP}.
Fig.~\ref{fig:overview}\subref{fig:proposed} shows the proposed network architecture.
The encoder uses the $1\times C$ SFI convolutional layer with the kernel size of $K$ and the stride of $S$.
The decoder is the $C\times 1$ SFI transposed convolutional layer with the kernel size of $K$ and the stride of $S$.
It is applied to $\bm{v}\odot \bm{v}_m$ for each $m$.
Although we can use a different SFI transposed convolutional layer for each $m$, we experimentally observed that using the same SFI transposed convolutional layer for all $m$ provided a higher separation performance.
Thus, we adopted this configuration for the decoder.

The architecture of the mask predictor is the same as SuDoRM-RF and depends on the input SF.
Thus, we use a method of adjusting $K$ and $S$ in accordance with the input SF \cite{Saito2022IEEEACMTASLP}.
It adequately changes the values of $K$ and $S$ to keep the time resolution of the pseudo time-frequency representation unchanged in units of second.
For example, when we trained the network with $K=240$ (5 ms) and $S=120$ (2.5 ms) at $F_s=48$~kHz, we set $K=40$ (5 ms) and $S=20$ (2.5 ms) at $F_s=8$~kHz.
In equation, 
\begin{equation}
      K^{\target}=\dfrac{F^{\target}_s}{F^{\train}_s}K^{\train},
      \quad
      S^{\target}=\dfrac{F^{\target}_s}{F^{\train}_s}S^{\train},
      \label{eq:adjust}
\end{equation}
where we use the superscripts ${}^{\train}$ and ${}^{\target}$ to denote the values for the training and test data, respectively.
When either $K^{\target}$ or $S^{\target}$ becomes non-integer, we can use the algorithms of handling noninteger kernel sizes and strides for the SFI layers \cite{KImamura202309EUSIPCO}.
Thus, we can ensure that the entire network is SFI.

In summary, the proposed network works as follows:
Given the input mixture, (i) it first generates the weights of the SFI layers using the input SF, (ii) then adjusts the kernel sizes and strides of the SFI layers in accordance with Eq.~\eqref{eq:adjust}, and (iii) finally separates the mixture into $M$ output signals.
Steps (i) and (ii) can be omitted after the first weight generation whenever the input SF is kept unchanged.

As described in Section~\ref{sec:sfi_layer}, the SFI layers have the same computational cost as their usual counterparts, except for the first weight generation in the test stage.
Thus, we can construct the proposed SFI extension of SuDoRM-RF without sacrificing the computational efficiency, one of the advantages of SuDoRM-RF.
In addition, we can use the loss function defined in Eq.~\eqref{eq:loss} for the training because the SFI layers do not require any special constraints on the loss function.
This allows the proposed network to inherit the capability of handling a variable number of sources.

\begin{table}[t]
\centering
{
\footnotesize
\caption{Number of training, validation, and test data in FUSS48k dataset}
\begin{tabular}{c|cccc|c}
      \toprule
      Split & $N=1$ & $N=2$ & $N=3$ & $N=4$ & Total \\ \midrule
      Training & 4992 & 4893 & 5072 & 5053 & 20000 \\
      Validation & 254 & 253 & 229 & 264 & 1000 \\
      Test & 237 & 249 & 262 & 252 & 1000 \\
      \bottomrule
\end{tabular}
\label{tab:dataset_spec}
}
\end{table}
\begin{figure*}[t]
      \centering
      \includegraphics[width=1.1\columnwidth]{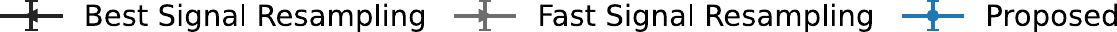}
      \vspace{1mm}
      \\
      \subfloat[$N=1$]{
            \includegraphics[width=0.5\columnwidth,clip]{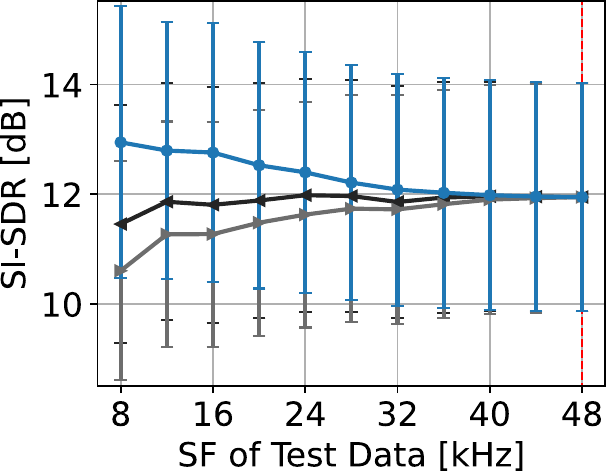}
      }
      \subfloat[$N=2$]{
            \includegraphics[width=0.5\columnwidth,clip]{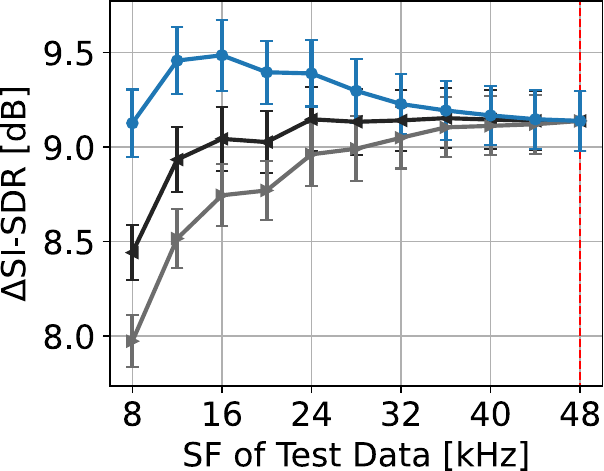}
      }
      \subfloat[$N=3$]{
            \includegraphics[width=0.49\columnwidth,clip]{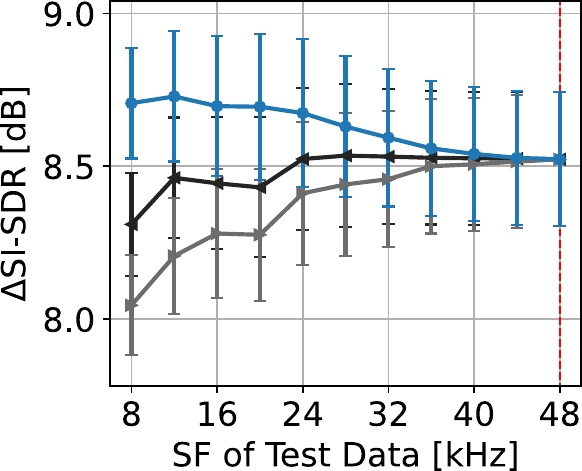}
      }
      \subfloat[$N=4$]{
            \includegraphics[width=0.5\columnwidth,clip]{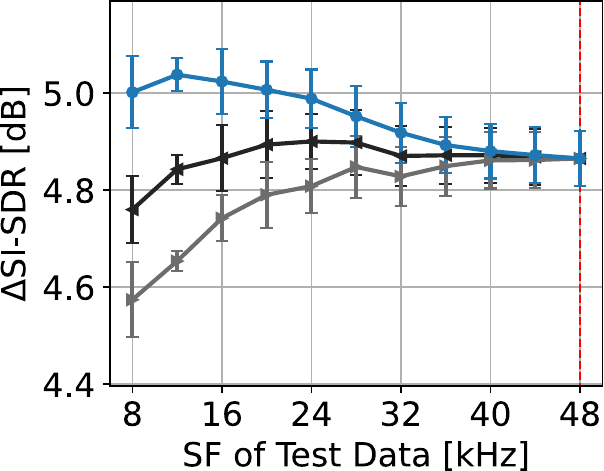}
      }
      \caption{
            Separation performances of proposed and signal-resampling-based methods for each $N$.
            Red dotted lines denote trained SF. Error bars show standard errors.
      }
      \label{fig:results}
\end{figure*}

\section{Experiments} \label{sec:exp}
\subsection{Experimental Conditions} \label{sec:exp_cond}
To evaluate the effectiveness of the proposed method, we conducted a USS experiment with a variable number of sources.

\pgsec{Dataset} The free universal sound separation (FUSS) dataset \cite{Wisdom2021ICASSP} has been popularly used in the USS studies, but the SF of the audio signals is too low (16 kHz) for the evaluation at a wide range of SFs.
Thus, we created a 48 kHz-sampled version of the FUSS dataset, which we call the \emph{FUSS48k dataset}.

We synthesized mixture signals of the FUSS48k dataset by modifying the implementation in the official repository of the FUSS dataset\footnote{\url{https://github.com/google-research/sound-separation/blob/master/datasets/fuss}}.
We set the SF to 48 kHz and the duration of each mixture to 8 s.
The maximum number of sources was set to four and the number of sources $N$ varied from one to four.
Table~\ref{tab:dataset_spec} shows the number of the mixtures for training, validation and test data.
After the synthesis, we resampled the test data at 8, 12, $\ldots$, 44 kHz, which were used for the evaluation as well as the 48 kHz-sampled test data.

\pgsec{Network} We used the proposed network with $M=4$ for all methods\footnote{
      Since the MGF has only three parameters and the generated weights have much less degrees of freedom, it may limit the separation performance compared with the original SuDoRM-RF.
      We should confirm its impact on separation performance, although the scope of this paper is not to achieve the state-of-the-art separation performance.
      Thus, we trained the original SuDoRM-RF by setting $K=240$ and $S=120$, which were the closest even values to the values used for the FUSS dataset in \cite{Tzinis2022JSPS}.
      However, all elements of the output signals of SuDoRM-RF were zeros after the first epoch in the training.
      Although we trained the network with several different values of $K$ and $S$, this numerical instability was not resolved.
      We concluded that the original SuDoRM-RF did not work well for 48-kHz sampled data and we did not use it for the comparison.
}.
For the SFI layers, we set $K=240,S=120,$ and $I=960$.
As the latent analog filters, we used a modulated Gaussian function\cite{Saito2022IEEEACMTASLP}:
\begin{equation}
      G(\omega;\mu,\sigma,\varphi)=e^{-(\omega-\mu)^2/(2\sigma^2)+ \j \varphi} + e^{-(\omega+\mu)^2/(2\sigma^2)- \j \varphi}, 
\end{equation}
where $\mu$ is the center angular frequency, $\sigma^2$ is the variance of the Gaussian, and $\varphi$ is the initial phase.
These parameters were defined by each input and output channel pair.
They were initialized as in \cite{Saito2022IEEEACMTASLP}, but $\sigma^2$ was initialized with $(50\pi)^2$.
The other parameters were the same as SuDoRM-RF ($C=512$ and $B=16$).

We used the training setup as \cite{Tzinis2022JSPS}.
The loss function was given as Eq.~\eqref{eq:loss}.
The optimizer was Adam with an initial learning rate of $1.0\times10^{-3}$, which was multiplied by $1/3$ every 10 epochs.
The gradient clipping was applied so that the maximum of the $l_2$ norms of the gradient was five.
The batch size was four and the number of epochs was 150.
As a data augmentation, we shuffled the source signals between samples in a minibatch and multiplied each source by a random gain.

\pgsec{Compared methods} The proposed method (\emph{Proposed}) was compared with two signal-resampling-based methods used in \cite{Saito2022IEEEACMTASLP}.
The signal-resampling-based methods resample the input mixture at 48~kHz, apply the trained model to the resampled mixture, and resample the output signals back to the input SF.
For signal resampling, we used two different resampling methods that prioritized resampling accuracy and fast computation, respectively.
\emph{Best Signal Resampling} used the accurate but slow method and \emph{Fast Signal Resampling} used the fast but less accurate method\footnote{
      As in \cite{Saito2022IEEEACMTASLP}, we used the \texttt{resample} function in the \texttt{librosa} library \cite{librosa}. The \texttt{res\_type} argument of this function was set to \texttt{kaiser\_best} for Best Signal Resampling and \texttt{kaiser\_fast} for Fast Signal Resampling.
}.
We stress again that all the methods used the same trained model and differ only in handling untrained SFs.

\pgsec{Evaluation metric} We used the scale-invariant signal-to-distortion ratio (SI-SDR) for $N=1$ and the SI-SDR improvement ($\Delta$SI-SDR) for $N=2,3,$ and 4.
To reduce the dependency of the parameter initialization, we trained the network with four different random seeds and computed the averages and standard errors of the metrics.

\subsection{Results}
\label{sec:results}

Fig.~\ref{fig:results} shows the separation performances of all methods.
The standard errors for $N=1$ were greater than those for the other $N$.
This should be because the difficulty of reconstructing the source signal depends on the source types.
As the SF decreased, the performance of the signal-resampling-based methods became lower for all $N$.
Fast Signal Resampling provided lower SI-SDRs and $\Delta$SI-SDRs than Best Signal Resampling.
The gap in these metrics between the two methods increased as the SF decreased.
These results show that the signal resampling (especially with lower resampling accuracy) can degrade the USS performance.

The proposed method provided comparable and higher performances on average than the signal-resampling methods for all $N$.
The improvement of Proposed from Best Signal Resampling increased as the test SF moved away from the trained SF.
Although the performance gap between Proposed and Best Signal Resampling tends to be lower for the greater $N$, this should be caused by the fact that the separation generally becomes more difficult as the number of sources increases.
This result shows that the SFI layers and the adjustment method of $K$ and $S$ are effective for USS.

\section{Conclusion}
We proposed a USS method capable of handling various SFs by applying the SFI layers to SuDoRM-RF.
The SFI layers generate the convolutional kernels in accordance with the input SF, which enables the network to handle various SFs.
Once the convolutional kernels are generated, these layers work as their usual counterparts whenever the input SF is kept unchanged.
Thus, the SFI layer can install the SF universality into SuDoRM-RF without sacrificing the universality with respect to source types.
While the proposed network contains the non-SFI subnetwork, it can be entirely SFI by adequately adjusting the kernel sizes and strides of the SFI layers in accordance with the input SF.
Experiments demonstrated that the signal resampling can degrade the USS performance and the proposed method can handle untrained SFs more effectively than the signal-resampling-based methods.
We believe that the proposed method is an important step toward the realization of the \emph{literally universal} sound separation.

\vfill\pagebreak

\bibliographystyle{IEEEbib}
\bibliography{refs}

\end{document}